\documentclass[11pt,twocolumn,twoside,a4paper,amsmath,amssymb,aps,showkeys,showpacs]{revtex4}

\usepackage{amsfonts}
\usepackage{graphics} 
\usepackage{epsfig}
\usepackage{fancyheadings}

\newcommand{\Fi}[1]   {Fig.~\ref{#1}}

\newcommand{\agev}    {\mbox{$A$~GeV}}               

\newcommand{\rb}[1]   {\mbox{\textrm{\scriptsize #1}}}
\newcommand{\rbt}[1]  {\mbox{\textrm{\tiny #1}}}
\newcommand{\sqrts}   {\ensuremath{\sqrt{s_{_{\rbt{NN}}}}}}

\newcommand{\lam}     {\ensuremath{\Lambda}}
\newcommand{\lab}     {\ensuremath{\bar{\Lambda}}}  
\newcommand{\xim}     {\ensuremath{\Xi^{-}}}

\newcommand{\xip}     {\ensuremath{\bar{\Xi}^{+}}}

\newcommand{\kmin}    {\ensuremath{\textrm{K}^-}}
\newcommand{\kplus}   {\ensuremath{\textrm{K}^+}}

\newcommand{\ommin}   {\ensuremath{\Omega^-}}
\newcommand{\omplus}  {\ensuremath{\bar{\Omega}^+}}           

\newcommand{\mtavg}   {\ensuremath{\langle m_{\rb{t}} \rangle - m_{\rb{0}}}}

\newcommand{\dndy}    {\ensuremath{\textrm{d}N/\textrm{d}y}}

\newcommand{\nwound}  {\ensuremath{\langle N_{\rb{w}} \rangle}}

\newcommand{\mub}     {\ensuremath{\mu_{\rbt{B}}}}
\newcommand{\gams}    {\ensuremath{\gamma_{\rb{s}}}}

\newcommand{\piavg}   {\ensuremath{\langle \pi \rangle}}

\newcommand{\kstar}   {\ensuremath{\textrm{K}^{*}\textrm{(892)}}}
\newcommand{\kstarbar}{\ensuremath{\bar{\textrm{K}}^{*}\textrm{(892)}}}

\begin{document}

\thispagestyle{myheadings}
\rhead[]{}
\lhead[]{}
\chead[C.~Blume]{Strangeness Production at the SPS}

\title{Strangeness Production at the SPS}

\author{C.~Blume, for the NA49 Collaboration}
\email{blume@ikf.uni-frankfurt.de}
\affiliation{
Fachbereich Physik, 
J.W.~Goethe-Universit\"at,  
Max-von-Laue-Str.~1,
D-60438 Frankfurt am Main,
GERMANY }

\received{}

\begin{abstract}
Systematic studies on the production of strange hyperons and the $\phi$~meson 
as a function of beam energy and system size performed by the NA49 collaboration
are discussed.  Hadronic transport models fail to describe the production 
of multi strange particles ($\Xi$, $\Omega$), while statistical models are 
generally in good agreement to the measured particle yields at all energies.  
The system size dependence is well reproduced by the core-corona approach.  
New data on \kstar\ production are presented.  The yields of these short-lived 
resonances are significantly below the statistical model expectation.  
This is in line with the interpretation that the measurable yields are 
reduced due to rescattering of their decay products inside the fireball.
\end{abstract}



\maketitle


\section{Introduction}
\label{introduction}

The production of strange particles has always been a key observable in 
heavy-ion reactions and its enhancement was one of the first suggested 
signatures for quark-gluon plasma (QGP) formation \cite{RAFELSKI}.  The 
predicted enhancement of strangeness production in nucleus--nucleus 
collisions relative to proton--proton reactions was established 
experimentally some time ago \cite{NA35LAM,NA35STR} and it was also 
found that this enhancement is increasing with the strangeness content 
of the particle type \cite{NA57HY158}.  However, a clear interpretation 
of these phenomena requires a systematic investigation of the energy and 
system size dependence of strangeness production.  In the following we 
report on some aspects of such a study done by the NA49 experiment.

\section{Energy Dependence}
\label{energy_dep}

%
\begin{figure}[t]
\includegraphics[width=0.95\linewidth]{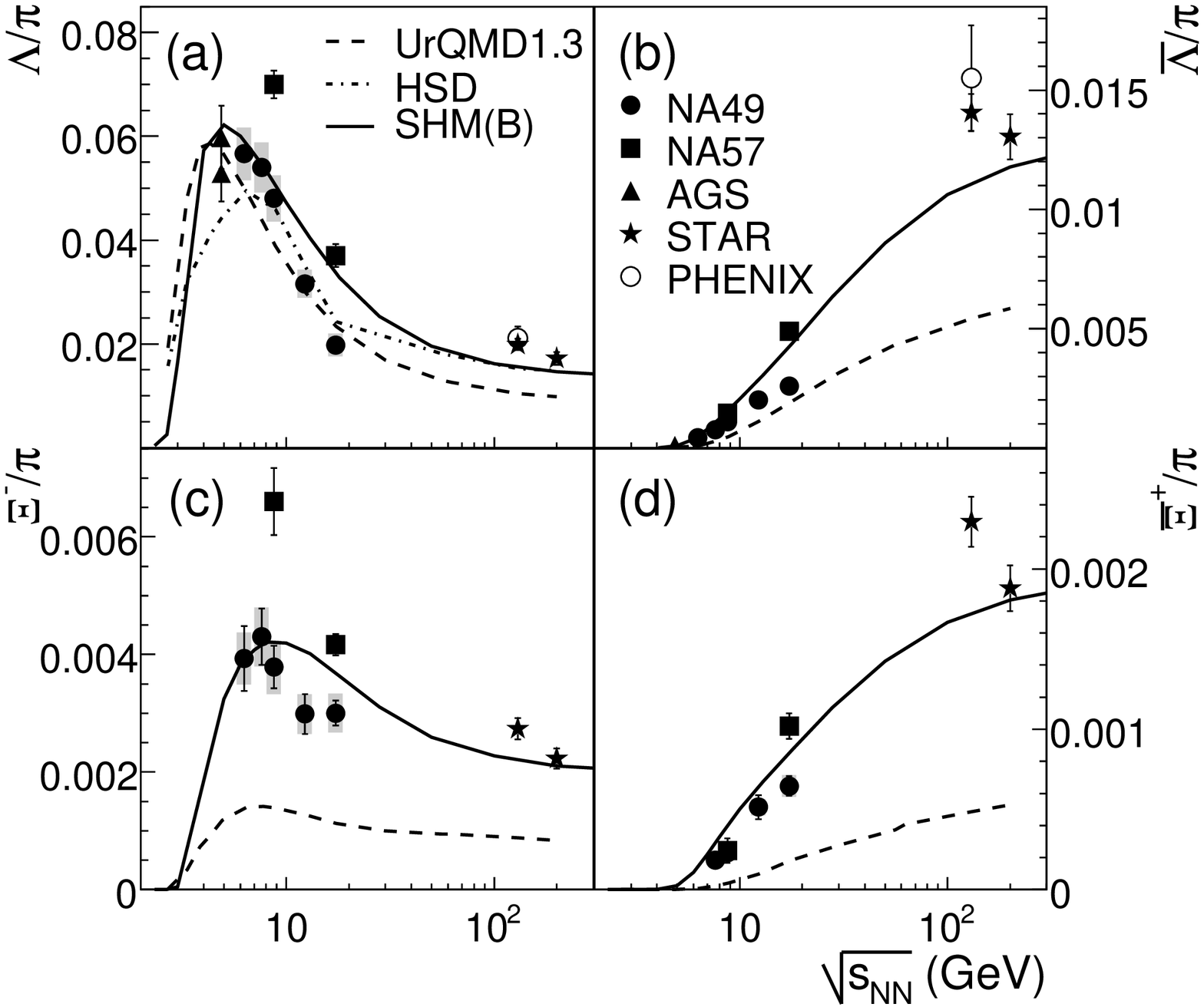}
\caption{The rapidity densities \dndy\ at mid-rapidity of \lam\ (a), \lab\ (b), 
\xim\ (c), and \xip\ (d) divided by the pion rapidity densities
($\pi = 1.5 \: (\pi^{+} + \pi^{-})$) in central Pb+Pb and Au+Au collisions as 
a function of \sqrts\ \cite{NA49HYEDEP}.  The systematic errors are
represented by the gray boxes. Also shown are NA57 \cite{NA57HY40,NA57HY158}, 
AGS \cite{E891LAM,E896LAM,E917LAB,E802PION}, 
and RHIC 
\cite{STARLM130,STARXI130,STARHY200,STARPI130,STARPI200,PHNXLM130,PHNXPI130} 
data, as well as calculations with hadronic transport models (HSD, UrQMD1.3 
\cite{HSD,URQMD,HSDURQMD}) and a statistical hadron gas model 
(SHM(B)~\cite{ANTON}).}
\label{fig:dndypion} 
\end{figure} 
%

\begin{figure}[t]
\includegraphics[width=0.85\linewidth]{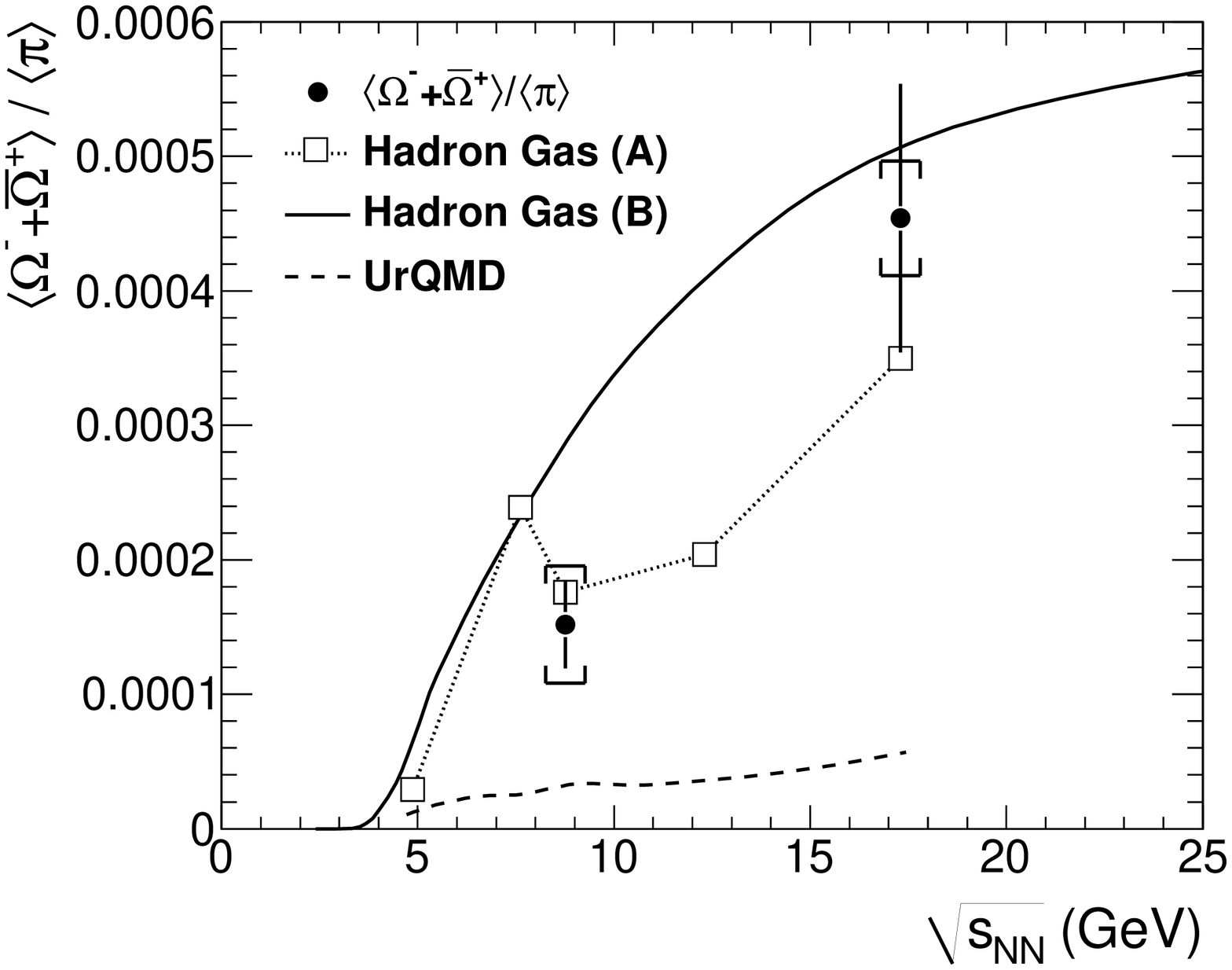}
\caption{\label{fig:omegaedep} 
The total yield of $\ommin + \omplus$ divided by the total number of
pions $\piavg$ ($\piavg = 1.5 \: (\pi^{+} + \pi^{-})$) versus the 
center-of-mass energy \cite{NA49OMEGA}.  The dashed curve shows the 
prediction from the hadronic transport model UrQMD1.3~\cite{URQMD}. 
A hadron gas model without strangeness suppression~\cite{REDLICH} is 
shown by the full curve.  The open squares represent the fits from 
\cite{BECATTINI} including a strangeness under-saturation factor \gams.}
\end{figure}

\begin{figure}[t]
\includegraphics[width=0.85\linewidth]{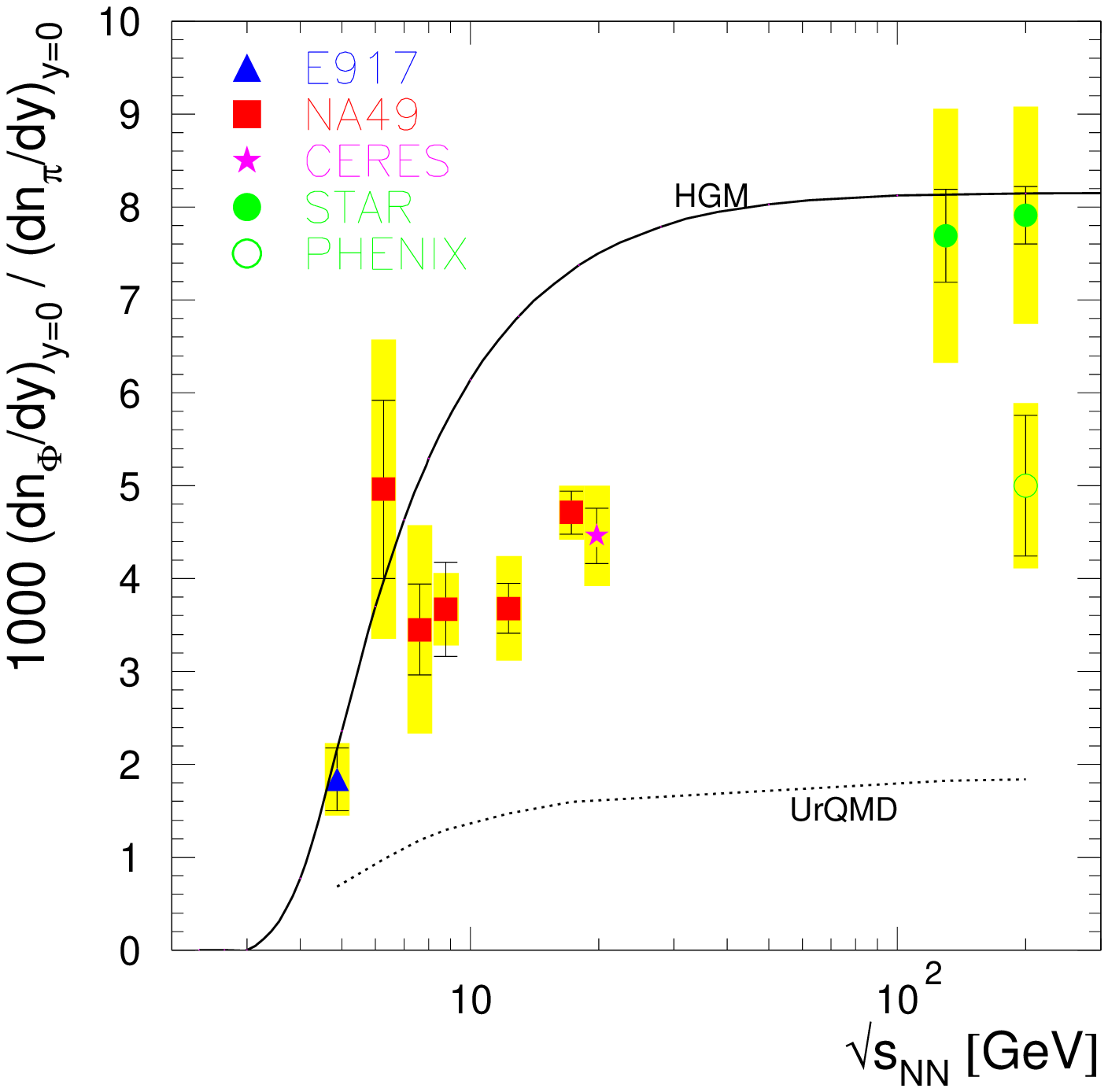}
\caption{\label{fig:phiedep} 
The rapidity densities \dndy\ at mid-rapidity of $\phi$ divided by the 
pion rapidity densities ($\pi = 1.5 \: (\pi^{+} + \pi^{-})$) in
central nucleus-nucleus collisions as a function of \sqrts\ \cite{NA49PHEDEP}.
Also shown are NA45/CERES \cite{NA45PHI}, and RHIC \cite{STARPHI,PHNXPHI} 
data, as well as calculations with hadronic transport models (UrQMD1.3 
\cite{URQMD}) and a statistical hadron gas model (HGM~\cite{ANTON}).}
\end{figure}

Figure~\ref{fig:dndypion} shows a comparison of the energy dependence of
mid-rapidity \lam, \lab, \xim, and \xip\ production to several models and 
results from other experiments.  While the transport models UrQMD1.3 and 
HSD provide a reasonable description of the \lam/$\pi$ amd \lab/$\pi$ 
ratios, they are clearly below the data points in case of the \xim\ and 
\xip.  This might indicate that an additional partonic contribution is 
necessary to reach the production rates observed for multi-strange 
particles.  Statistical models on the other hand generally provide a 
better match to the data.  These models are based on the assumption that 
the particle yields correspond to their chemical equilibrium value and 
can thus be described by the parameters temperature $T$, baryonic chemical 
potential \mub, volume $V$, and, in some implementations, by an additional 
strangeness under-saturation factor \gams. The curves shown in 
\Fi{fig:dndypion} labeled SHM(B) are taken from \cite{ANTON} and are based 
on parametrizations of the \sqrts\ dependence of $T$ and \mub.  

The difference between the two model approaches discussed here is even 
more prominent for the $\Omega$, as demonstrated in \Fi{fig:omegaedep}.  
In this case the deviation to the hadronic transport model is of the order
of a factor of 10, while both the statistical model approaches shown in 
\Fi{fig:omegaedep} are quite close to the data points.

While multi-strange hyperons generally seem to be close to the full 
equilibrium expectation at all energies, the $\phi$-meson exhibits significant 
discrepancies (see \Fi{fig:phiedep}).  While at lower energies the 
$\phi$ production is close to both, the statistical model and the transport 
model UrQMD1.3, at top SPS energies none of the models does match the 
measurements.  Please note that the appearant discrepancy of UrQMD1.3 with 
the $\phi$/$\pi$~ratios at lower energies, as visible in \Fi{fig:phiedep}, 
is rather due to an overestimate of the pion yields and not an 
underestimate of the $\phi$ yields \cite{NA49PHEDEP}.  Also shown in 
\Fi{fig:phiedep} is a measurement of the $\phi$ yields via the dielectron 
decay $\phi \rightarrow \textrm{e}^{+} + \textrm{e}^{-}$ performed by the 
NA45 collaboration at 158\agev\ \cite{NA45PHI}.  This result agrees quite 
well with the NA49 result, which has been measured using the hadronic decay 
branch $\phi \rightarrow \textrm{K}^{+} + \textrm{K}^{-}$.

\section{System Size Dependence}
\label{size_dep}

%
\begin{figure}[t]
\begin{center}
\includegraphics[width=\linewidth]{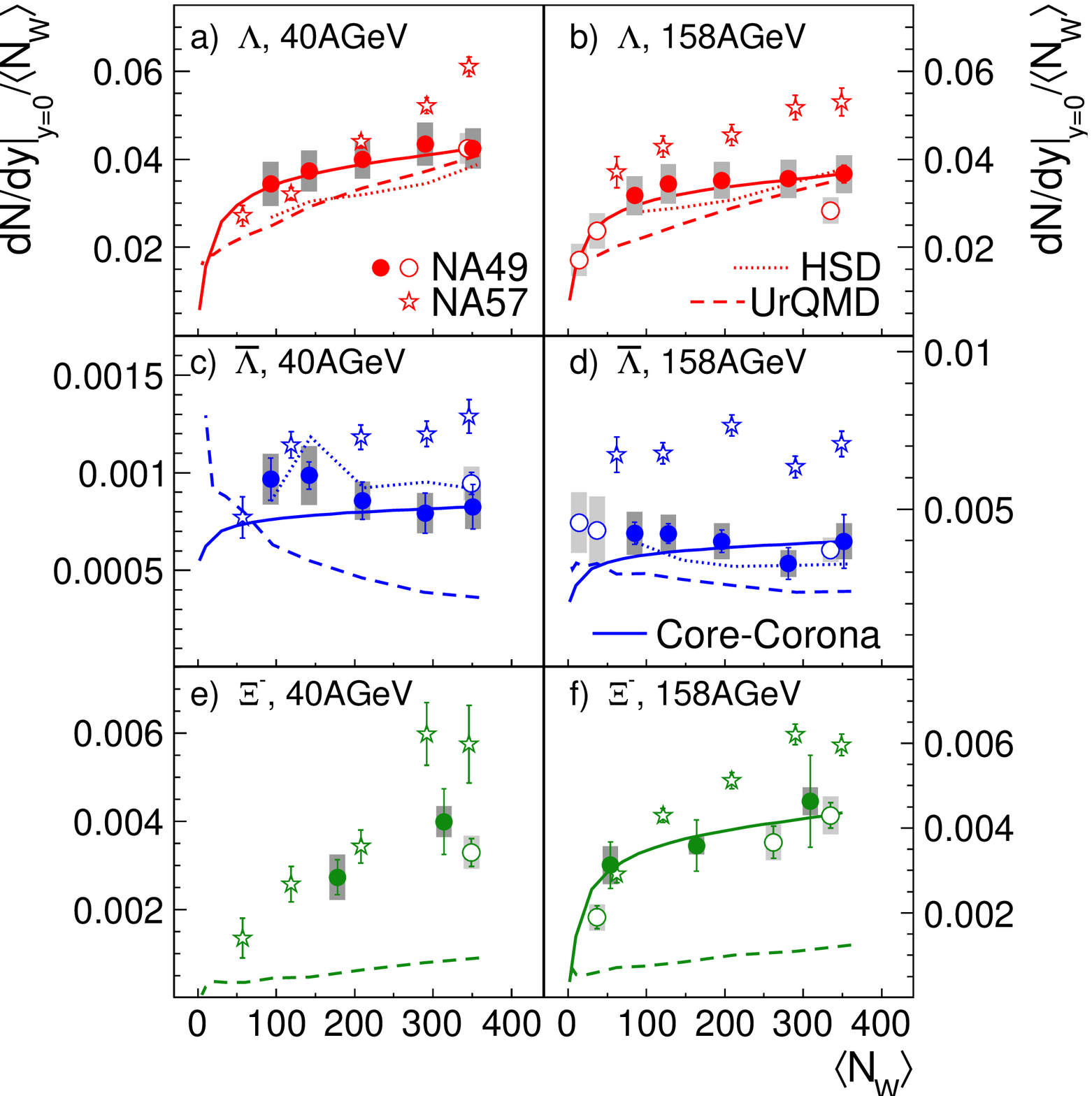}
\end{center}
\caption{
The rapidity densities \dndy\ divided by the average number of wounded 
nucleons \nwound\ of \lam, \lab, and \xim\ at mid-rapidity for Pb+Pb 
collisions at 40$A$ and 158\agev, as well as for near-central C+C and 
Si+Si reactions at 158\agev, as a function of \nwound\ \cite{NA49HYSDEP}.   
Also shown are data of the NA57 collaboration \cite{NA57HY40,NA57HY158} 
(open stars) and calculations with the HSD model \cite{HSD} 
(dotted lines), the UrQMD2.3 model \cite{URQMD,URQMD23} (dashed 
lines), and the core-corona approach (solid lines) \cite{BECATTINI2,WERNER}.
}
\label{fig:dndy_vs_nw}
\end{figure}
%

%
\begin{figure}[t]
\begin{center}
\includegraphics[width=0.85\linewidth]{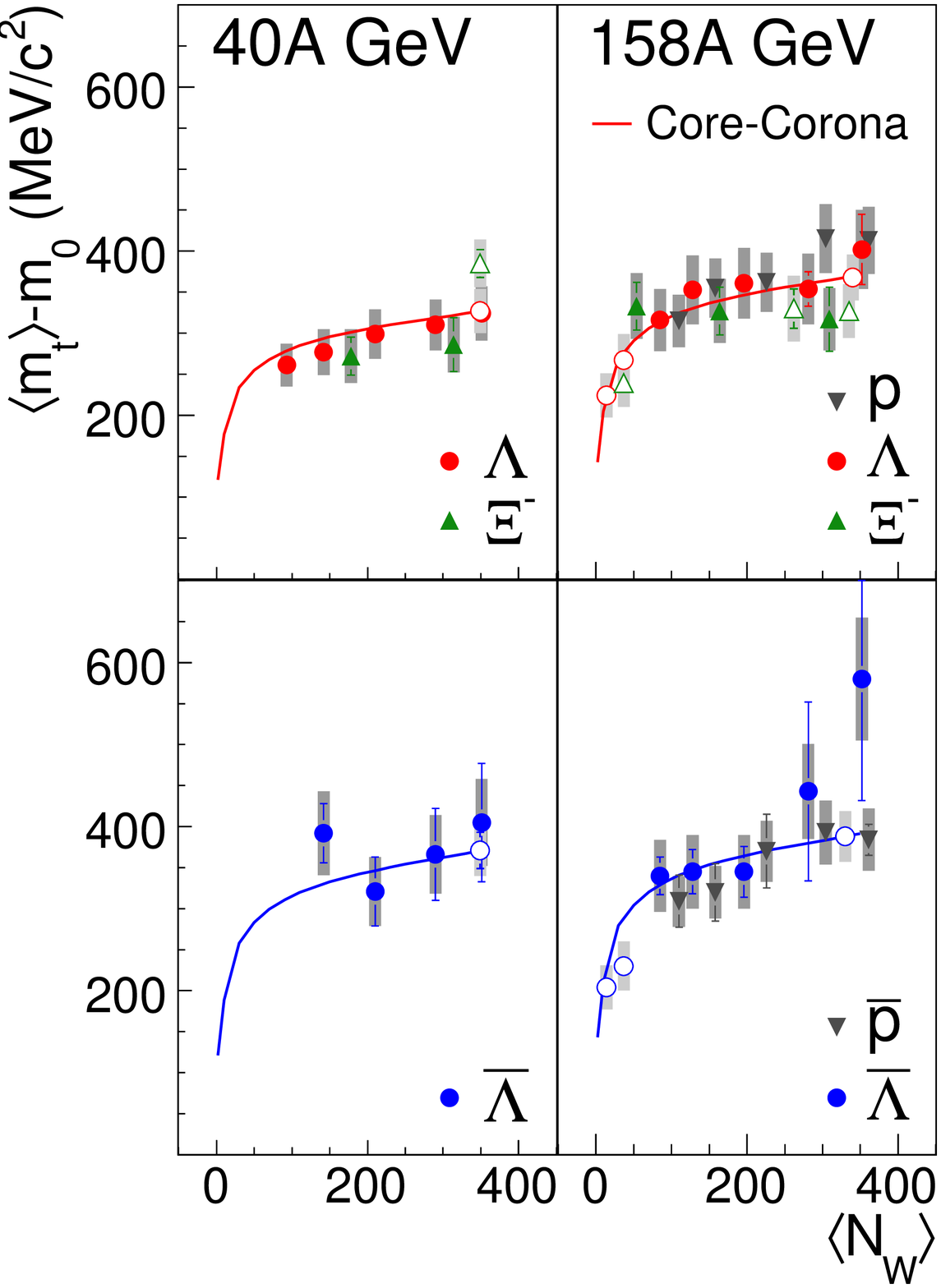}
\end{center}
\caption{
The \mtavg\ values at mid-rapidity for Pb+Pb collisions at 40$A$ 
and 158\agev, as well as for near-central C+C and Si+Si reactions at 
158\agev\ \cite{NA49HYSDEP}.  The (anti-)proton data are taken from 
\cite{NA49SDPR1}.  Also shown are the results from a fit for \lam\ and 
\lab\ with the core-corona approach (solid lines).}
\label{fig:meanmt_vs_nw} 
\end{figure} 
%

The system size dependence of \lam, \lab, and \xim\ production close to
mid-rapidity, as measured at SPS energies, is summarized in 
\Fi{fig:dndy_vs_nw}.  For \lam\ and \lab\ a relatively early saturation 
at $\nwound \approx 60$ is observed by NA49.   However, a clear discrepancy
between the data of NA49 and NA57 is still present.  The transport models 
UrQMD2.3 \cite{URQMD23} and HSD \cite{HSD} are close to the data points 
for \lam, but are slightly below the \lab\ measurements.  The 
\xim\ production is clearly under-predicted at all system sizes.  The 
core-corona approach \cite{BECATTINI2,WERNER} provides generally a much 
better description of the system size dependence of all strange particle 
species.  Here the relevant quantity is the fraction of nucleons that 
scatter more than once $f(\nwound)$ which can be calculated in a Glauber
model.  This allows for an interpolation between the yields $Y$ measured 
in elementary p+p ($= Y_{\textrm{corona}}$) and in central nucleus-nucleus 
collisions ($= Y_{\textrm{core}}$):
\begin{eqnarray*}
Y(\nwound) & =  \nwound & \: [ f(\nwound) \: Y_{\textrm{core}} \\ 
           &            & \: + \: (1 - f(\nwound)) \: Y_{\textrm{corona}} ]
\end{eqnarray*}
Please note that the curves shown in \Fi{fig:dndy_vs_nw} and 
\Fi{fig:meanmt_vs_nw} are based on a function $f(\nwound)$ that was 
calculated for Pb+Pb interactions.  Therefore their comparison to the 
smaller systems C+C and Si+Si is not directly possible, since their
surface to volume ratio is different.  

It is interesting to observe that this approach not only works for
yields, but also for dynamical quantities such as \mtavg\ (see 
\Fi{fig:meanmt_vs_nw}).  This suggests that the core-corona picture 
provides in general a reasonable way for understanding the evolution
from elementary p+p to central Pb+Pb collisions.

\section{Resonances}
\label{resonances}

%
\begin{figure}[t]
\begin{center}
\includegraphics[width=0.85\linewidth]{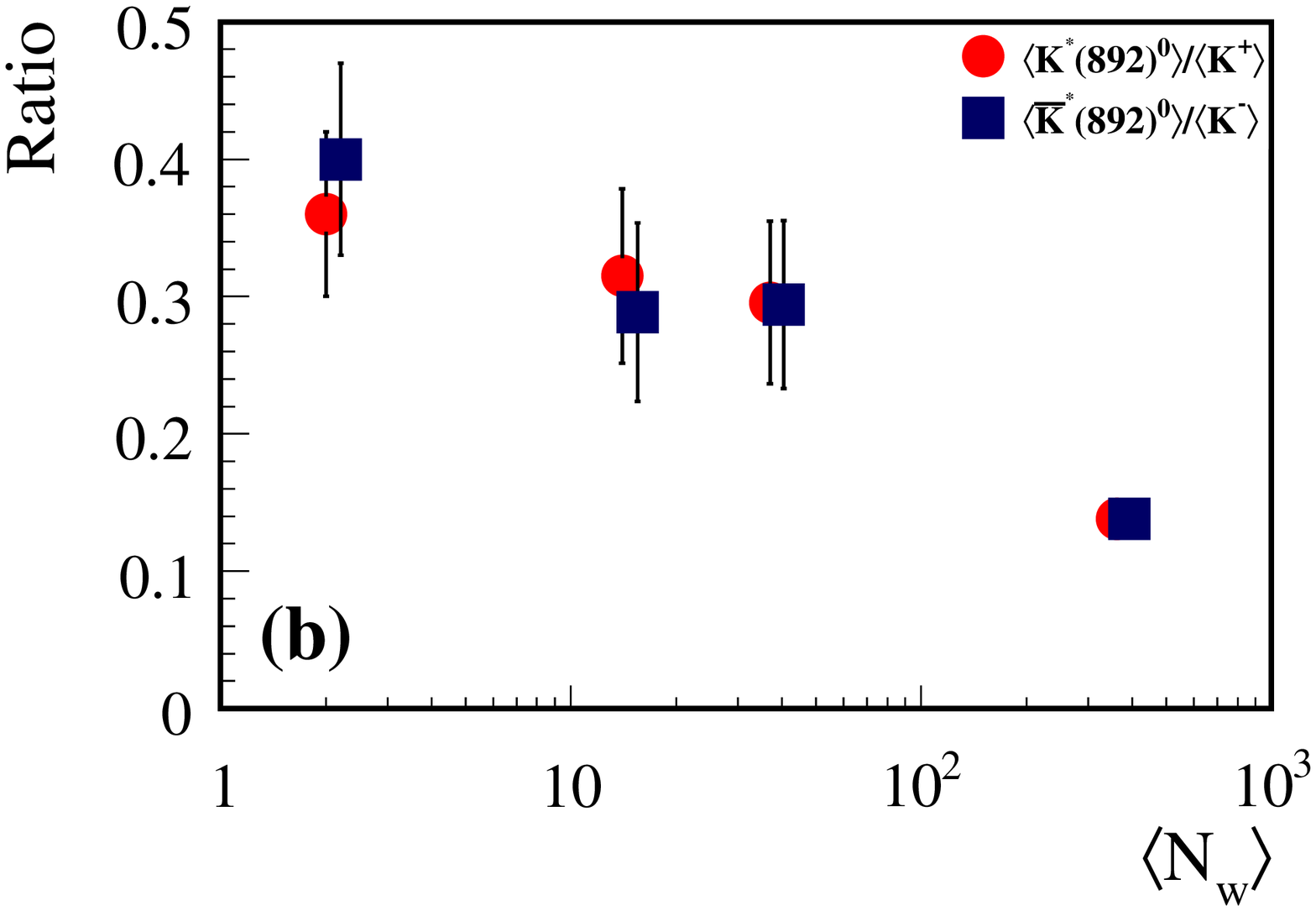}
\end{center}
\begin{center}
\includegraphics[width=0.88\linewidth]{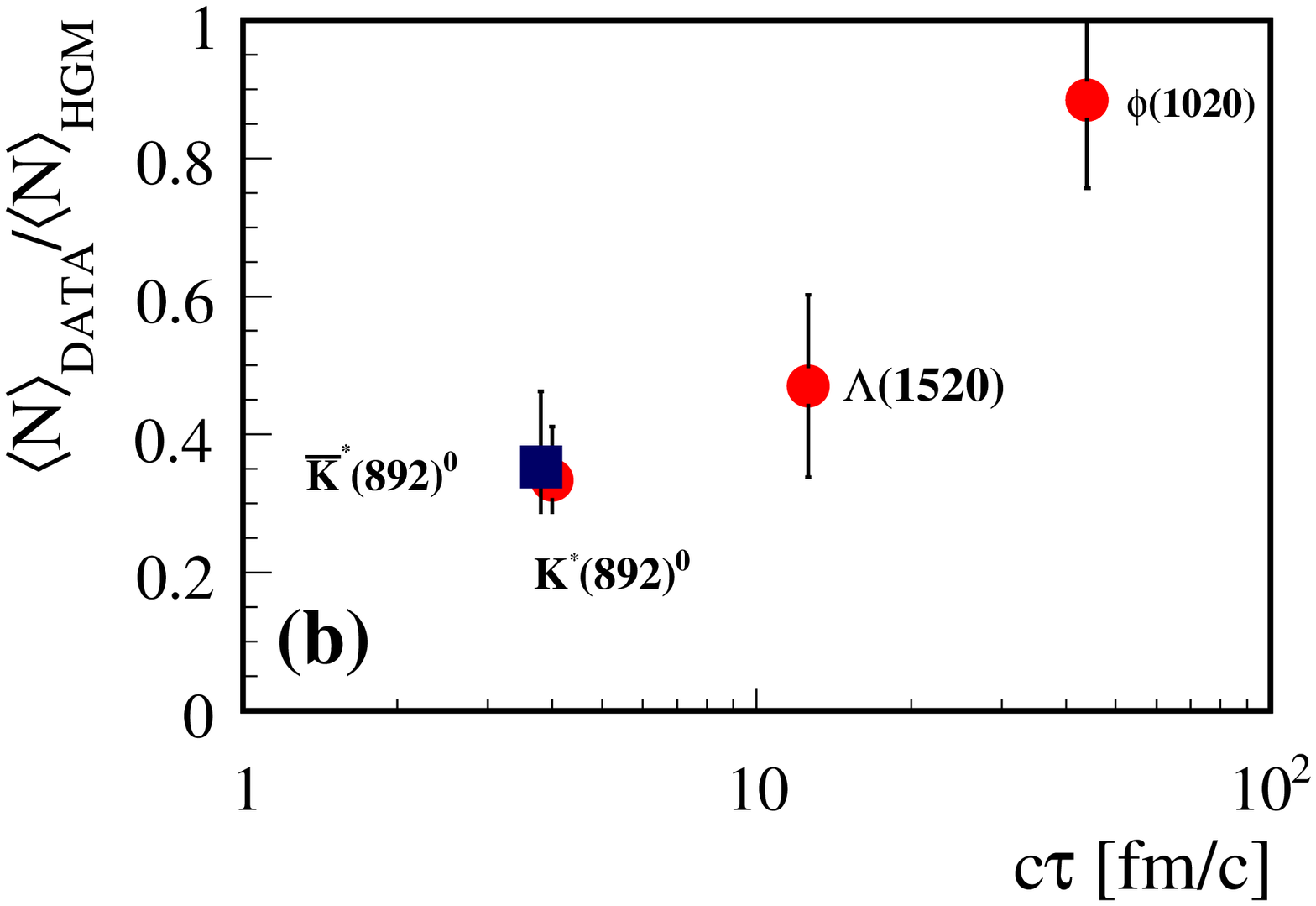}
\end{center}
\caption{
Upper panel: The total yield of \kstar\ (\kstarbar) divided by
the total yields of \kplus\ (\kmin) in p+p and nucleus-nucleus
collisions at 158\agev\ as a function of the average number of 
wounded nucleons \nwound.
Lower panel: The total yield of \kstar\ (\kstarbar),
$\Lambda\textrm{(1520)}$, and $\phi$ in central Pb+Pb collisions
at 158\agev\ divided by the expectation from a statistical model
fit \cite{BECATTINI3} as a function of the resonance lifetime $c\tau$.
}
\label{fig:kstar} 
\end{figure} 
%

Strange resonances are of particular interest due to their short lifetimes
that are in the same order as the lifetime of the fireball.  Because of
this their yields can still be modified after chemical freeze-out via
destruction and regeneration mechanisms.  For instance the particles
resulting from the decay of such a resonance can rescatter in the fireball
such that the resonance cannot be reconstructed any more.  These effects 
can thus lead to deviations from the chemical equilibrium expectation.  

New data on the \kstar\ (\kstarbar) production in central 
nucleus-nucleus collisions at 158\agev\ are summarized in \Fi{fig:kstar}.  
The \kstar\ (\kstarbar) are reconstructed via the decay 
$\kstar \rightarrow \kplus + \pi^{-}$ ($\kstarbar \rightarrow \kmin + \pi^{+}$).
As shown in the upper panel of \Fi{fig:kstar}, the system size dependence of
the total \kstar\ yield is clearly different than the one of charged kaons, 
the ratios \kstar/\kplus\ (\kstarbar/\kmin) decrease with increasing system 
size.  This could be indicative of a stronger reduction of the measurable 
\kstar\ yields in the larger fireball of central Pb+Pb reactions compared
to the smaller one produced in C+C and Si+Si collsions, because here their 
decay products have a higher probability of rescattering with the medium.  

The lower panel of \Fi{fig:kstar} compares the total yields of several 
resonances (\kstar, $\Lambda\textrm{(1520)}$, and $\phi$) to the expectations 
from a statistical model fit \cite{BECATTINI3}.  The fit did not include
the resonances themselves.  The deviation is largest for the short lived 
\kstar, while it is slighly less pronounced for the $\Lambda\textrm{(1520)}$ 
and even less for the $\phi$, which has a much longer lifetime than the other 
two resonances.  Comparing the yields of resonances with different lifetimes 
can thus provide a means to study the time-like extension of the hot and
dense fireball created in heavy ion reactions.

\label{last}

\end{document}